\documentclass[
reprint,
superscriptaddress,
showpacs,preprintnumbers,
 amsmath,amssymb,
]{revtex4-1}
\usepackage{graphicx}
\usepackage{dcolumn}
\usepackage{bm}
\usepackage{color}

\begin{document}

\preprint{APS/123-QED}

\title{Electron Surfing and Drift Accelerations	 in a Weibel-dominated High-Mach-number Shock}

\author{Yosuke Matsumoto}
\email{ymatumot@chiba-u.jp}
\affiliation{%
Department of Physics, Chiba University, 1-33 Yayoi-cho, Inage-ku, Chiba 263-8522, Japan}%
\altaffiliation[Also at ]{Institute for Global Prominent Research, Chiba University, 1-33 Yayoi-cho, Inage-ku, Chiba 263-8522, Japan}%


\author{Takanobu Amano}
\affiliation{
 Department of Earth and Planetary Science, the University of Tokyo, 7-3-1 Hongo, Bunkyo-ku, Tokyo 113-0033, Japan}%

\author{Tsunehiko N. Kato}
\affiliation{Center for Computational Astrophysics, National Astronomical Observatory of Japan, 2-21-1, Osawa, Mitaka, Tokyo 181-8588, Japan}%

\author{Masahiro Hoshino}
\affiliation{
 Department of Earth and Planetary Science, the University of Tokyo, 7-3-1 Hongo, Bunkyo-ku, Tokyo 113-0033, Japan}%


\date{\today}

\begin{abstract}
How electrons get accelerated to relativistic energies in a high-Mach-number quasi-perpendicular shock is presented by means of ab initio particle-in-cell simulations in three dimensions. We found that coherent electrostatic Buneman waves and ion-Weibel magnetic turbulence coexist in a strong-shock structure whereby particles gain energy during shock-surfing and subsequent stochastic drift accelerations. Energetic electrons that initially experienced the surfing acceleration undergo pitch-angle diffusion by interacting with magnetic turbulence and continuous acceleration during confinement in the shock transition region. The ion-Weibel turbulence is the key to the efficient nonthermal electron acceleration.
\end{abstract}

\pacs{52.35.Tc, 52.65.Rr, 96.50.Pw, 98.70.Sa}
\maketitle

Elucidating the acceleration mechanisms of charged particles have been of great interests in laboratory, space, and astrophysical plasma physics. Among other mechanisms, a collisionless shock is thought to be an efficient particle accelerator. This idea has been strengthened by radio, X-ray, and gamma-ray observations of astrophysical objects such as supernova remnant shocks, indicating that protons and electrons are efficiently accelerated to TeV energies at such very strong shock waves \citep{Reynolds1986,Bamba2003,Aharonian2004,Ackermann2013}. Efficient electron acceleration at high-Mach-number shocks was also recently suggested by in-situ measurements at Saturn's bow shock \citep{Masters2013}. Motivated by these circumstances, laboratory experiments using high-power laser facilities have emerged to provide a new platform for tackling such problems \citep{Morita2010,Kuramitsu2011,Ross2012,Ahmed2013,Meinecke2014,Park2015}.

The diffusive shock acceleration (DSA) theory \citep{Bell1978,Blandford1978} has provided a solution to observational evidences for efficient accelerations at collisionless shocks, as it predicts a power-law energy spectrum of particles having a spectral index that is close to the values suggested by multi-wavelength observations. As the DSA theory presumes pre-existing mildly energetic particles, pre-acceleration mechanisms are required to provide a seed population for DSA, particularly for electrons \citep{Amano2007,Kang2014}. The connection between pre-acceleration and DSA remains a critical issue in shock acceleration theory.

One possible pre-acceleration mechanism is the so-called shock drift acceleration (SDA), in which a particle gains energy in the shock transition region during its gradient-$|\bm{B}|$ drift. For an electron with a Larmor radius much smaller than the shock thickness, the interaction time with the shock (and hence the energy gain) is determined by the adiabatic theory \citep{Wu1984,Leroy1984,Krauss-Varban1989}. Subsequent acceleration can be realized by self-generated electromagnetic waves excited by accelerated electrons \citep{Amano2010,Matsukiyo2011,Guo2014a,Kato2015,Park2015a}. 

Alternatively, the shock surfing acceleration (SSA) becomes particularly important for electrons in high-Mach-number perpendicular shocks \citep{Shimada2000,McClements2001,Hoshino2002,Schmitz2002}. This process uses large-amplitude electrostatic waves generated by the Buneman instability at the leading edge of the shock. Extensive numerical experiments have been performed to demonstrate its high efficiency of producing relativistic particles under sufficiently high-Mach-number conditions \citep{Amano2009,Matsumoto2012,Matsumoto2013,Wieland2016}. The generation of strong magnetic turbulence caused by the ion-Weibel instability in the shock transition region is also an important aspect in this regime \citep{Kato2010}. Electrons can be energized by spontaneous reconnection triggered throughout the region by the Weibel magnetic turbulence \citep{Matsumoto2015}.

Because these mechanisms have been proposed rather independently, how they interplay with each other remains unclear. Indeed, naively, it might be anticipated that the SSA would be killed by the strong magnetic turbulence. In this Letter, contrary to this expectation, we show prolonged electron acceleration in a high-Mach-number, quasi-perpendicular shock by means of three-dimensional (3D) particle-in-cell (PIC) simulations. The present {\it numerical shock experiments} enabled us to investigate for the first time the physics of electron acceleration in a fully self-consistent fashion that included all of the essential ingredients: SDA, SSA, and a strong Weibel magnetic turbulence.

We used a SIMD-optimized, hybrid-parallel 3D PIC simulation code, which implements the quadratic particle weighting function for accurate treatment of particle accelerations with a limited number of particles per cell \citep{Cormier-Michel2008}. A shock wave was created by the so-called injection method, in which particles are continuously injected from one side of the simulation boundary (here, $x=L_x$) at supersonic (super Alfv\'enic) speed $V_0$ in the $-x$-direction and specularly reflected at the other side of the simulation boundary ($x=0$). The shock front propagates in the $+x$-direction in the present shock-downstream frame. We adopted the simulation parameters of the ion-to-electron mass ratio of $M/m=64$ and the upstream plasma $\beta=1$, which is equally shared by ions and electrons ($\beta_i=\beta_e=0.5$). The upstream magnetic field has the $x$- and $z$-components $\bm{B}_0 = (B_{0x},0,B_{0z})$, such that the shock angle becomes $\tan^{-1}({B_{0z}/B_{0x}})=\Theta_{B_n}=74.3^\circ$ and the upstream motional electric field has only the $y$-component as $E_{0y} = -V_0B_{0z}/c$ with $c$ being the speed of light. The resulting sonic (Alfv\'en) Mach number reached $M_s\sim22.8$ ($M_A\sim20.8$) with a non-relativistic upstream velocity of $V_{up}/c\sim0.26$ measured in the shock-rest frame. Thus, the generated shockwave falls into the sub-luminal shock where $\tan^{-1}(c/V_{up}) > \Theta_{B_n}$. The electron inertia length $c/\omega_{pe}$ is resolved with 20 computational cells, and one numerical time step resolves $0.025\omega_{pe}^{-1}$, where $\omega_{pe}$ is the electron plasma frequency in the upstream region. The simulation domain size in the $x$-direction ($L_x$) expands as the shockwave propagates, while the shock front spans 4.8 times the ion inertia length ($\lambda_i$) wide in the $y$- and $z$-directions. We discuss space and time in units of the ion inertia length and inverse the ion gyro frequency ($\Omega_{gi}^{-1}$) in the upstream region, and particles' momentum and energy in the shock-rest frame. 20 particles per cell per species were used in the upstream region. In total, one trillion particle motions were followed in the simulation domain with $8800\times768\times768$ computational cells in the latest time development. Such computationally demanding simulations were made possible by using 9216 nodes (73,728 processor cores) and 100 TB of physical memory on the Japanese K computer.

\begin{figure}
\includegraphics[width=0.475\textwidth]{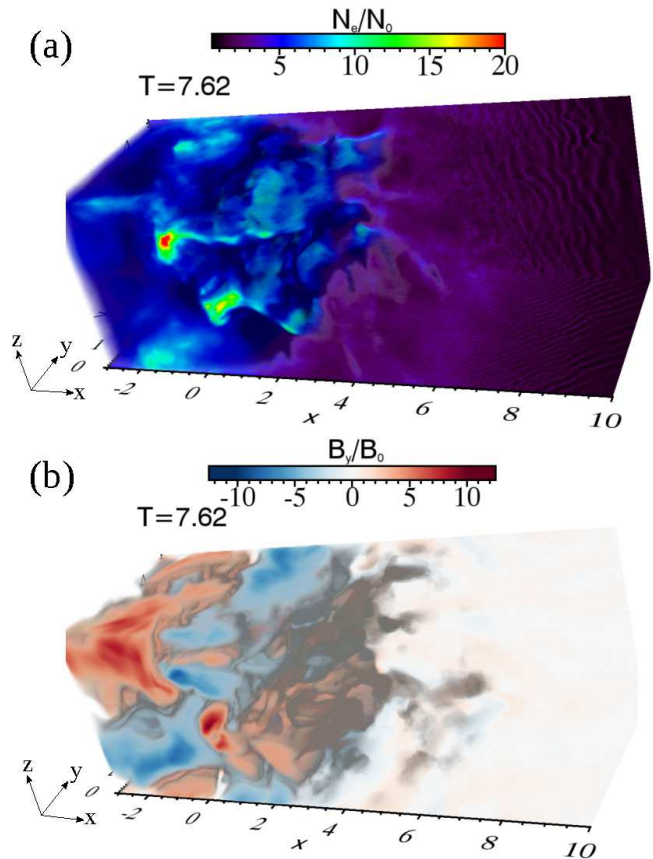}
\caption{The 3D structure around the shock front ($x=0$) obtained at time $T=7.62$ from the initiation of the experiment. The structures of (a) the electron density and (b) the $y$-component of the magnetic field are visualized by a volume rendering technique with cross-sectional profiles in the $x$-$y$ ($z=0$) and $x$-$z$ ($y=L_y$) planes. The quantities and the spatial scale were normalized to the upstream values and upstream ion inertia length, respectively. Videos of time evolution corresponding to (a) and (b) are provided as Supplemental Material.}
\label{fig:ov3d}
\end{figure}

Fig. \ref{fig:ov3d} shows a 3D shock structure in a fully developed stage after initiation. Electron-scale coherent structures were found to persist during the simulation run, as can be seen from the stripes of electron density at the leading edge of the shock ($8<x<10$) in Fig. \ref{fig:ov3d}(a). The shock transition (foot) region ($0<x<6$) was dominated by the ion-Weibel instability because of an interaction between the upstream and reflected ions, resulting in rib structures (Fig. \ref{fig:ov3d}(a)) and strong magnetic turbulence (Fig. \ref{fig:ov3d}(b)). The $y$-component of the magnetic field is a component newly generated by the instability and is further amplified up to $20$ times the upstream value by the shock compression, as are the other $x$- and $z$-components.

\begin{figure}
\includegraphics[width=0.475\textwidth]{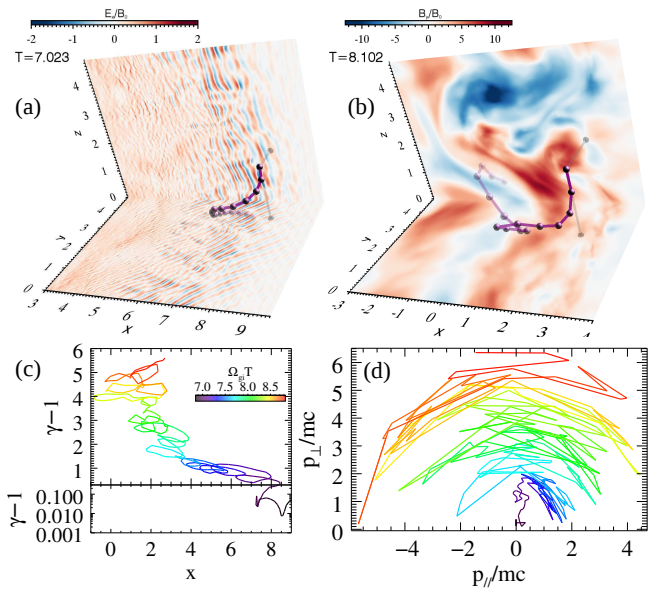}
\caption{(a) The selected electron's position at $T=7.0$ is shown with a magenta sphere with grey lines and spheres projected onto the profiles of the $x$-component of the electric field in the $x$-$y$ ($z=0$) and $x$-$z$ ($y=L_y$) planes. The trajectory back in $7.5\ \Omega_{ge}^{-1}$, where $\Omega_{ge}$ is the upstream electron gyro frequency, is expressed by magenta spheres with a gradual decrease in opacity. (b) Electron trajectory at $T=8.1$ is in the same format as (a), but with the $y$-component of the magnetic field profiles. (c) Particle energy (Lorentz factor $\gamma-1$) history as a function of distance in the $x$-direction from the shock front location. The color of the line corresponds to the time shown in the color bar. (d) Time history of the particle's momentum with respect to the local magnetic field direction. Videos of the electron's orbit corresponding to (a) and (b) are provided as Supplemental Material.}
\label{fig:accel}
\end{figure}

To understand how electrons are accelerated by interacting with such coherent and turbulent structures, we selected about $10^7$ tracer particles self-consistently solved in the PIC simulations. They initially shared the same $x$ coordinates within a cell width in the upstream region at time $T=6.8$ in the fully developed stage. The particle's motion was recorded every $5\ \omega_{pe}^{-1}$ until the majority were transmitted downstream at $T=8.8$. The time histories of position, energy, and momentum of the most energetic electron in the final time of tracking are presented in Fig. \ref{fig:accel}. 

At the leading edge, electron-scale, coherent electrostatic waves are excited with amplitudes of $|E| > B_0$ (Fig. \ref{fig:accel}(a)), as is also seen in the electron density profile. The Buneman instability is driven unstable because of the interaction between the upstream electrons and the reflected ions in this region to produce the large-amplitude waves. The wave front is oblique to the $x$-axis in the $x$-$y$ plane, reflecting the gyrating motion of the reflected ion. There is no characteristic structure in the $z$-direction, indicating that the most unstable mode lies in the two-dimensional (2D) plane. The selected electron orbit in Fig. \ref{fig:accel}(a) showed an abrupt change in motion when it entered this Buneman-destabilized region. The particle was then accelerated in the direction opposite to the motional electric field in the $y$-direction while being trapped by the electrostatic wave front. This picture is essentially the same as the electron SSA in 2D as previously reported \citep{Amano2009,Matsumoto2012,Matsumoto2013}, which is surprising because the coherent potential structure persists even in the 3D system in which the Buneman instability can excite, in general, many oblique modes. Note also that the Buneman and ion-Weibel instabilities can coexist in different regions. This allows the coherent SSA to operate virtually without any interference from the Weibel magnetic turbulence.

After experiencing the SSA, the particle can be further energized by interacting with the turbulent fields around the shock front. As the particle's gyro radius becomes comparable to the size of the magnetic irregularities with a typical scale of the ion inertia length, the pre-accelerated electron does not follow simple $\bm{E}\times\bm{B}$ or gradient-$|\bm{B}|$ drift motions. Rather, it undergoes strong pitch-angle scattering by the magnetic turbulence (Fig. \ref{fig:accel}(b)). 

Fig. \ref{fig:accel}(c) shows the energy (Lorentz factor $\gamma$) history of the particle. The Lorentz factor initially increased with $\Delta \gamma \sim 1$ during the SSA at the leading edge ($7 < x < 8.5$), followed by convective transport towards the shock front. As the particle approaches the shock front, the energy again increases  continuously while being diffused spatially in the $x$-direction around the shock front. The time evolution in the momentum space in Fig. \ref{fig:accel}(d) characterizes aspects of these acceleration processes. During $6.9<T<7.3$, the momentum increases preferentially in a direction perpendicular to the magnetic field, reflecting direct acceleration by the motional DC electric field during the SSA. The particle is then scattered in the momentum space. The pitch-angle scattering becomes stronger as the particle approaches the shock front, and the momentum increases in both the parallel and perpendicular directions in $7.5<T<8.8$.

\begin{figure}
\includegraphics[width=0.475\textwidth]{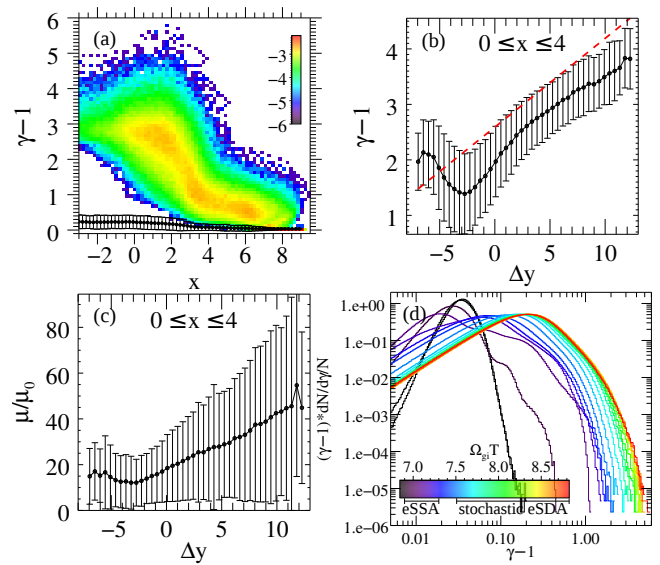}
\caption{(a) Probability distribution in the phase space (position in $x$ and energy) for the most energetic particles during tracking in $6.8<T<8.8$. The probability is color-coded on a logarithmic scale. The average values for the thermal particles are plotted with error bars representing the standard deviation in each bin. (b) The average energy of the energetic particles in $0 \leq x \leq 4$ as a function of displacement in the $y$-direction from each particle's initial position in the upstream region ($\Delta y$). The energy increase rate from the motional electric field is indicated by the red dashed line. (c) The first adiabatic invariant of the energetic particles in $0 \leq x \leq 4$ as a function of $\Delta y$ normalized to the upstream value using the bulk speed $V_{up}$ and the magnetic field strength $B_0$. (d) Time evolution of the tracer particles' energy distribution. The colors of the lines correspond to times in the color bar.}
\label{fig:pdf}
\end{figure}

Here we take a statistical approach to understand the overall acceleration history. We selected a population consisting of the top 1000 highest-energy electrons at the final tracking time, termed the "most energetic particles". We also randomly sampled 1000 electrons from the rest of the tracer particles called "thermal particles". Next we analyzed the histories of those particles in phase space (position and velocity) during the tracking time. Fig. \ref{fig:pdf}(a) shows the probability distribution in the phase space for the most energetic particles. At first, the particles experienced an energization after passing the Buneman-destabilized region at the leading edge ($7 \le x \le9$), in contrast to a faint thermalization of the thermal particles (black line). 

The surfing-accelerated particles can be further accelerated as they approach the shock front. They attain energies mostly in the foot region in $0 \le x \le 4$, as shown by the particle's orbit in Fig. \ref{fig:accel}(c). During the second acceleration stage, the most energetic particles drift on average in the positive $y$-direction opposite to the motional electric field direction. Fig. \ref{fig:pdf}(b) shows the average energy of the energetic particles in $0 \le x \le 4$ as a function of each particle's displacement in the $y$-direction from the initial position in the upstream region ($\Delta y$). The energy monotonically increases as the particles drift in the $+y$-direction at a rate expected from the motional electric field $E_0$ and $\Delta y$, i.e., $\Delta \gamma = eE_{0y} \Delta y/mc^2$ (red dashed line), where $e$ is the elementary charge, showing that the energy gain itself is the same as the SDA. 

The major drawback of the SDA is the limited acceleration time determined by the adiabaticity of the particle's trajectory. A particle must escape from the shock transition region after a finite period of time, which terminates the acceleration process. On the other hand, in the presence of scattering, the stochasticity may allow the energetic particles to be confined and accelerated within the acceleration region much longer than expected according to the adiabatic theory. To check this hypothesis, we analyzed the history of the first adiabatic invariant $\mu = p_{\perp}^2/2m|\bm{B}_p|$ during the second acceleration stage, where $p_{\perp}$ is the particle's perpendicular momentum with respect to the local magnetic field direction $\bm{B}_p$, as shown in Fig. \ref{fig:pdf} (c). First, the invariant increases by a factor of $\sim15$ from the initial value via the SSA. Second, the invariant increases monotonically as the particles drift in the $+y$-direction in $0 \le x \le 4$. The acceleration process in this phase appears to be non-adiabatic, and the assumption made in the SDA theory is clearly violated. This suggests that the particle acceleration process can be understood as a stochastic SDA, in which the stochasticity is introduced by the Weibel magnetic turbulence along with the pre-acceleration by the SSA. (In contrast, the thermal particles were transmitted downstream quickly because the motion was adiabatic.)

The energy histories of all tracer particles are presented in Fig. \ref{fig:pdf} (d). The energy distribution, which was initially very cold, quickly deformed with a high-energy tail during $6.9 < T < 7.3$ as a result of the SSA. In the later time evolution, the entire distribution shifts to higher energy until $T\sim 8.1$. After that, position of the thermal peak stays at $(\gamma-1) \sim 0.2$. The high-energy tail beyond this peak extends to form a power-law distribution slowly but continuously with its cut-off energy at $(\gamma-1) \sim 5$ at the final tracking time of $T\sim 8.8$.

\begin{figure}
\includegraphics[width=0.475\textwidth]{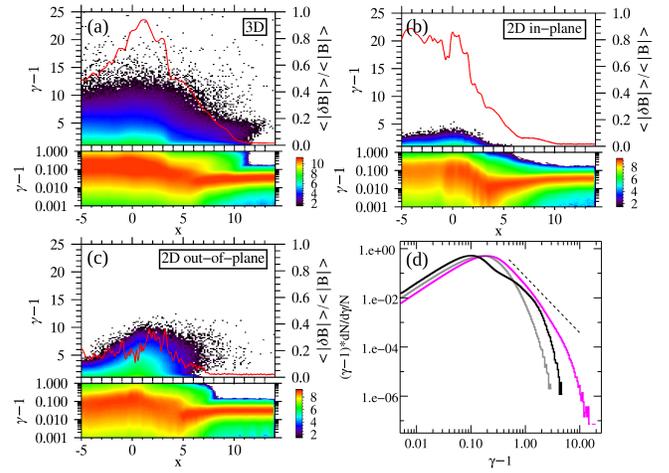}
\caption{(a) Electron phase space density in the upstream and downstream regions at $T=7.8$. The color and the ordinate axis of the lower energy part ($\gamma-1 < 1$) are in logarithmic scales. Normalized magnetic field fluctuation $\langle|\delta \bm{B}|\rangle / \langle |\bm{B}| \rangle$ is also plotted with the red solid line referring the right ordinate axis. Here $\langle \rangle$ denotes the average in the $y$--$z$ plane and $\delta \bm{B} = \bm{B}-\langle\bm{B}\rangle$. (b) and (c) Results from 2D simulations for the in-plane and out-of-plane upstream magnetic field cases, respectively. (d) Energy distributions in the downstream region ($-5.0<x<-4.5$) from the 3D (magenta line), 2D in-plane (gray line), and 2D out-of-plane (black solid line) runs. The black dashed line indicates a power-law distribution with the index of -3.5.}
\label{fig:edist}
\end{figure}

Fig. \ref{fig:edist}(a) shows a snapshot of the phase space density of electrons sampled in the upstream and downstream regions at $T=7.8$, which includes particles experiencing various acceleration time histories. Very high energy electrons with $\gamma>5$ diffused into the upstream region and were even found in the far upstream regions ($x>10$). The diffusion was accomplished by strong magnetic fluctuations as quantified by $\delta B / B$ in Fig. \ref{fig:edist} (a) (red solid line); the turbulence level around the shock front reached almost the mean magnetic field strength ($\delta B / B \sim 1$).

The systems lacking either the SSA or the Weibel turbulence could not lead to production of such very high energy particles as complemented by 2D simulations under the same upstream conditions. The in-plane upstream magnetic field case, in which the Buneman mode was weakly destabilized \citep{Riquelme2011,Wieland2016}, resulted a faint heating in the turbulent area (Fig. \ref{fig:edist}(b)). On the other hand, the efficient SSA was realized in the out-of-plane case (Fig. \ref{fig:edist} (c)). The surfing-accelerated electrons, however, cannot undergo the subsequent stochastic SDA because of the weak turbulence level ($\delta B/B \sim 0.3$), which is due to limited growth of the ion Weibel instability in the 2D out-of-plane upstream magnetic field condition \citep{Wieland2016}.

The downstream energy spectrum from the 3D case is shown in Fig. \ref{fig:edist}(d) (magenta line), the lower-energy part ($(\gamma-1) < 1$) of which is well represented by the tracer particles' distribution (Fig. \ref{fig:pdf}(d)). The difference in cut-off energy between the tracer particles and the whole particle distribution (representing longer time history) again implied efficient confinement of the energetic particles in the acceleration region, leading to the formation of a power-law energy spectrum with the index of $-3.5$. This contrasts with the downstream Maxwell distribution in the 2D in-plane case (gray solid line) and with small amount of nonthermal particles in the 2D out-of-plane case (black solid line).

Simulation runs for super-luminal cases (i.e., $\tan^{-1}(c/V_{up}) < \Theta_{B_n}$) also resulted limited acceleration efficiency as previously reported for magnetized relativistic shocks \citep{Sironi2013}. Therefore, the present mechanism works essentially in non-relativistic quasi-perpendicular shocks, but with wide range of the upstream magnetic field obliquity provided $c/V_{up} \gg 1$. Another mechanism associated with magnetic reconnection in the Weibel turbulence \citep{Matsumoto2015} was not observed here because the Mach number was not large enough (as also confirmed by the 2D simulations in Fig. \ref{fig:edist}(b)). Nevertheless, this process may also come into play at even higher Mach-number shocks to boost the overall acceleration efficiency. Following the time evolution much further will eventually illuminate a whole process including self-excited low-frequency electromagnetic waves in the upstream region by the relativistic electrons \citep{Amano2010} and their subsequent participation in the DSA cycle in high-Mach-number quasi-perpendicular shocks.

\acknowledgments
This research used the computational resources of the K computer provided by the RIKEN Advanced Institute for Computational Science through the HPCI System Research project (Project ID: hp150133, hp150263, hp160212, hp170231) and Cray XC30 at Center for Computational Astrophysics, National Astronomical Observatory of Japan, and  was supported in part by JSPS KAKENHI  Grant Number 26400266, and MEXT as “Priority Issue on Post-K computer” (Elucidation of the Fundamental Laws and Evolution of the Universe) and JICFuS.


\end{document}